\documentclass[epsfig,a4paper,12pt]{article}
\usepackage{epsfig} 
\usepackage{amsmath}

\begin{document} 
{\thispagestyle{empty}
\setcounter{page}{1}
\title{Hysteretic optimization for the Sherrington--Kirkpatrick spin glass}
\author{K\'aroly F.~P\'al} 
 
\maketitle
\centerline{Institute of Nuclear Research of the Hungarian Academy of Sciences} 
\centerline{H-4001 Debrecen, P.O.~Box 51, Hungary, e-mail: kfpal@hal.atomki.hu}

\vspace*{0.21truein}
\begin{abstract}
Hysteretic optimization is a heuristic optimization method based on
the observation that magnetic samples are driven into a low energy state 
when demagnetized by an oscillating magnetic field of decreasing amplitude.
We show that hysteretic optimization is very good for finding ground states
of Sherrington--Kirkpatrick spin glass systems. With this method it is possible to
get good statistics for ground state energies for large samples of systems
consisting of up to about 2000 spins. The way we estimate error rates may be
useful for some other optimization methods as well. Our results show that
both the average and the width of the ground state energy distribution
converges faster with increasing size than expected from earlier studies. 
\end{abstract}
\centerline{PACS codes: 02.60.Pn, 07.05.Tp}
\centerline{Keywords: optimization, hysteresis, spin glass}

\section{Introduction}
Hysteretic optimization \cite{hystop} is based on the observation that
demagnetization of a magnetic sample with an oscillating magnetic field
of slowly decreasing amplitude leaves the sample in a very stable
low energy state. The most obvious application of the algorithm is
finding the ground state of models of disordered magnetic systems, which
is often a very hard optimization problem. In this case the method simply
consists of simulating the evolution of the system under the effect of the
appropriately varying field. The results can be further improved by
shaking up the system repeatedly. That means applying the demagnetization
procedure again and again, but with a maximum amplitude too small to align the
system fully with the field. The direction of the field is chosen randomly at
each site, and a different field pattern is used in each shake-up.
This strategy is better than doing full demagnetization cycles the same
number of times. A shake-up tends to preserve the best
correlations achieved so far, and it is also faster. Hysteretic
optimization is simulated demagnetization followed by shake-ups.

The algorithm may be generalized to a wide range of optimization problems
by extending the notion of the external field \cite{hystop, hyobookch}. That can be
done more than one way \cite{hystofs}. The performance of the generalized
algorithm has been demonstrated on instances of the travelling salesman
problem \cite{hystoptsp, hyobookch}. For every problem we have tried to solve so far with the
algorithm, some of its variations
significantly outperformed simulated annealing \cite{siman}, the
most popular general-purpose heuristic optimization algorithm.
However, in most cases there are other methods which are much better
suited for the particular problem. Hysteretic optimization
is not really effective for magnetic systems of low connectivity.
It does find very low lying states, but even our best attempts \cite{hystofs} were
unable to locate the true ground state of Edwards--Anderson spin glasses
of sizes that can be handled reliably and often even easily by some other algorithms
\cite{genalsgpal, genalsgpal2,genalsghart0, genalsghart, martinrg1, martinrg2,
extrop, middlet}. The method is also not effective for the random field Ising model
\cite{AlBaCo}.

For finding ground states of magnetic systems
of high connectivity, the situation is much better. For the
Sherrington--Kirkpatrick model the algorithm is very competitive.
We will show strong circumstancial evidence in the present paper
that it can find ground states of systems containing up to about
2000 spins reliably, in computation times that make affordable to treat
even a thousand of such samples on a few ordinary personal computers.
No algorithm has been  reported to show similar performance.
Boettcher determined ground states of 244 systems of size $N=1023$ \cite{BoeSK}},
as far as we know no such results for larger systems have been reported so far.
He used extremal optimization \cite{extrop}, and each case took almost a day
of computation. The present algorithm about two orders of magnitude faster.
It is also much more effective than hybrid genetic algorithm, which is very
good for the Edwards--Anderson case \cite{genalsgpal, genalsgpal2}.
Exact algorithm \cite{kobe} presently only affordable for quite small systems.

In the next section we discuss the algorithm. Then we explain how we determined
the reliability of the method for SK systems, we present our results, and
finally, we draw conclusions. 
\vspace*{-0.5pt}
\noindent

\section{The algorithm}
The energy of the Sherrington-Kirkpatrick Ising spin glass may be written as:
\begin{equation}
{\cal H}=-\frac{1}{2} \sum_{i,j}^N
J_{ij}\sigma_i\sigma_j-H\sum_i^N\xi_i\sigma_i.
\label{eq:sghami}
\end{equation}
The aim of the optimization problem is to find the spin values $\sigma_i=\pm 1$
that minimize the first sum. That configuration corresponds to the ground state
of the system at zero field. The $J_{ij}=z_{ij}/\sqrt{N}$ interactions
are fixed, random values, with $z_{ij}$ is either chosen according to a
Gaussian distribution with zero mean and unit variance (Gaussian case), or
they have values $+1$ or $-1$ with equal probability and independently from each
other ($\pm J$ case). Different choices of $J_{ij}$ define different instances
of the problem. Unlike in the case of the Edwards--Anderson model,
each spin interacts with all the others. The second sum is the external
demagnetizing field we apply to minimize the energy. The $H$ is the
appropriately oscillating field strength, while $\xi_i=\pm 1$, which
defines the direction of the field (together with the sign of $H$) at spin
site $i$, is chosen randomly for each $i$, and is fixed during each demagnetizing
cycle (either a full cycle or a shake-up). The size of the system is
characterized by $N$, the number of spins.

The full demagnetization process starts from the $\sigma_i=\xi_i$ state
aligned with the external field. This is the most favourable state
for a large positive field strength. At a certain field value, which can
easily be determined, it becomes favourable for a spin to flip.
The flip of that spin may destabilize other spins. When we simulate
the evolution of the system, we choose one of the unstable spins randomly, and
flip it. Then we determine the new list of favourable spin flips,
taking into account the effect of the flip on the stability of the other spins.
We repeat that until the avalanche stops, that is the system gets into a
stable state at that field strength. Then we decrease the field
further to the value when the next spin becomes unstable, and simulate
the avalanche its flip causes, while keeping $H$ fixed again. We keep
decreasing the field strength until we reach a negative value of
$H=-\gamma H_0$, while
following the evolution of the system from avalanche to avalanche.
This simulation corresponds to the limit of changing the field very slowly.
(For some problems it is better to change the field fast \cite{hystofs}.)
The $H_0$ is the value when the field is just strong enough to align the
system completely.
Its accurate determination is not at all critical, especially because we
use full demagnetization only as the initial phase before a series of
shake-ups. We have simply chosen $H_0=1.6$ for every instance. The $\gamma<1$
is the amplitude reduction factor, we have used $\gamma=0.95$. When we reach
the $H=-\gamma H_0$ field strength, we start increasing the field up
to $+\gamma^2H_0$, then decrease again to $-\gamma^3H_0$,
and so on. The demagnetization cycle is finished when the amplitude is
so small that there is no further spin flip. Every time we cross zero
field, we check the energy, and save the configuration whenever it is
better than the best one found before. If we are interested in the true
ground state, it is well worth doing it, because it often happens that
the final configuration of the cycle is not as good as something
encountered a few periods earlier. A shake-up is just another
demagnetization cycle, but it starts from the state the system was
left by the previous cycle, a state stable at zero field, and its maximum
amplitude is much smaller than $H_0$. We used $H_{shake}=0.7$ throughout
the present paper. Earlier we recommended starting each shake-up from the best
state found so far, which leads to a faster advance initially,
therefore it is a better strategy if we do a smaller number of
shake-ups. However, it makes somewhat harder to explore regions of the
configuration
space far from the current best state, which is disadvantageous in a long
run. Large systems do have very good configurations far from each
other. A very detailed description of the algorithm is given in
Ref.~\cite{hyobookch}.

In an algorithm outlined above, with repeated application of the same
procedure many times, it is an important question to decide when to stop.
To calculate reliable averages over random instances we need a large
sample for each system size to reduce statistical error, so we can not
spend too much time on a single case. At the same time, we want to find
the true ground state in a great majority of the samples to get small
systematic errors. The simplest possible stopping condition is to specify
the same, fixed number of shake-ups
$n_s$ for each instance of a given size $N$ (we will call
this stopping condition type 1). The problem with this prescription is that
it is hard to tell in advance how many shake-ups will be needed to get a
reasonable compromise between the two conflicting requirements above,
furthermore, this way we will
spend the same amount of time on the easiest and the hardest instances.
With our heuristic optimization method we have no way
to tell for any specific instance whether we have found the ground
state or not. However, if the algorithm is good enough
to find the ground state once, it must be able to find it again.
At the same time, if we are just slightly above the ground state energy, the
density of states the demagnetization process may
reach becomes so high that those states are hardly ever found more than once.
Our experience is that
there are not many states altogether that the algorithm finds
more than once in a reasonable time, and they are all among the lowest
lying states. If we find the current best state several times without
finding any better one, we may suppose that it is the ground
state. We can never be sure, but with a heuristic algorithm we can not
hope more than a low enough probability of failure. We will call stopping
condition type 2 to require finding the current lowest
energy in a prescribed number of shake-ups $M_{req}$, and accept it as the global
optimum. Similar termination conditions were used e.~g.\ in Refs.~\cite{lomast,bope2,payo},
and very probably also in many other papers, although such technical
details are not always stated explicitly. As a shake-up does not destroy
the configuration completely, finding a state in a shake-up somewhat
increases the probability of finding it again. If the maximum amplitude
is not too small, this is not a problem. We may have to specify a somewhat larger
$M_{req}$ to get the same reliability as we would get if out attempts were completely
independent from each other. We actually applied a mixed
terminating condition. Besides requiring an $M_{req}$ number of repetitions of the
current best energy, we also prescribed a minimum number of shake-ups
$n_{smin}$, so that we should not accept a suboptimal state too early,
and we stop if a maximum number $n_{smax}$ of shake-ups
is done to avoid using an excessive amount of time on the hardest
instances.

\section{Estimation of the reliability}

As we have mentioned before, whatever terminating condition we use, we can
never be sure we find the true ground state. If we have a series
of results on a large sample, we can hope to get a good order of
magnitude estimate of the probability of missing the
ground state by repeating the whole series again with the same stopping
condition, and checking how often the results differ. The run
giving the higher energy has certainly missed the ground state. We will estimate
the error rate of the algorithm with a given stopping condition
by counting how often a second series of runs
gives a worse result than the first one did. As we only need to count the cases
the second series surely missed, we may stop immediately if we reach an energy equal
or lower than the one the first run accepted as the ground state energy,
independently of the stopping condition to be tested. Full length calculation
is only needed for the hopefully very few cases that this series misses.
If the stopping condition is good enough to find the true ground state for
a great majority of instances, the first run certainly had to find it several
times in most cases. A repeated run stops at the first hit, so
the extra effort required to estimate the error rate this way is only
a fraction of what the original calculation needed.

A problem with this approach is that it almost certainly underestimates
the error rate. We register as missed cases
the ones whose ground state was missed by the repeated run, but found by the first
run, and the ones missed by both, with the second run doing worse. 
However, besides those, the correct estimate should also include
the cases when both runs gave the same suboptimal energy, and also
the ones for which the repeated run improved the energy, but still failed to find
the true ground state. We can not include these cases, because we
can not recognize them, not being able to
tell apart cases missed by both runs from other possibilities.
If all instances of the same size were equally
difficult, this were not a problem: the probability of failing in two
independent attempts were the square of the probability of failing the
first (or the second) time, so whenever we estimated a small enough
value, the error of this estimate would be negligible. 
However, some instances are orders of magnitude more difficult than others,
so we have no information about the proportion of cases that both runs missed.

To get some idea how much we underestimate the error rate this way,
we can do long calculations on not very large systems such that the estimated
error rate is extremely small. Then we can make two series of calculations
for the same instances with a much worse stopping condition, and estimate
the error rate as above, that is check how frequently the second series
gives a worse energy than the first one. As now we have the results
of the much more reliable long calculations as well, we can determine the true
error rate of our
calculation with the bad stopping condition, at least in a very good
approximation, and we can compare it to the estimated one. We can make this
comparison between estimated and true error rates for a few system sizes
with several different stopping conditions of a wide range of reliability. 
Actually, we need not do all those calculations with different terminating
conditions at all to derive these results. If we have enough
details of the run with the better condition, we can
tell what would happen if we applied a worse one, which would
stop the run earlier. In case of either type of stopping
conditions discussed in the present paper, it is enough to know
in which shake-up an improved energy was found the first time, and how many
times it was found before a further improvement happened.
For a type 1 stopping condition, we can easily tell what energy we
would end up if we stopped after any number of shake-ups, which is less than
what we actually did. If we applied a type 2 stopping condition,
we would accept a suboptimal state as ground state whenever we
found it the specified  $M_{req}$ number of times as the current best one.
In case of the mixed stopping condition, finding a suboptimal state
$M_{req}$ times only stops the run if no further improvent happens
before the minimum number of shake-ups is reached.
In the repeated runs we did not apply the original stopping condition at all, we only
stopped when we reached the energy we thought to be the optimum.
This way we could make estimates for the error rate for a variety
of stopping conditions from a single series for each system size.
We note that Martin \cite{martinbk} also suggested a self-consistent reliability test
for heuristic algorithms, but it does not seem to be readily applicable to
estimate the reliability with the terminating conditions we applied here.
For a low error rate a quantitative estimate that way would require
quite a long extra calculation, even with stopping condition type 1.

\section{Results}

\begin{figure}[ht]
\begin{center}
\scalebox{0.55}{\includegraphics{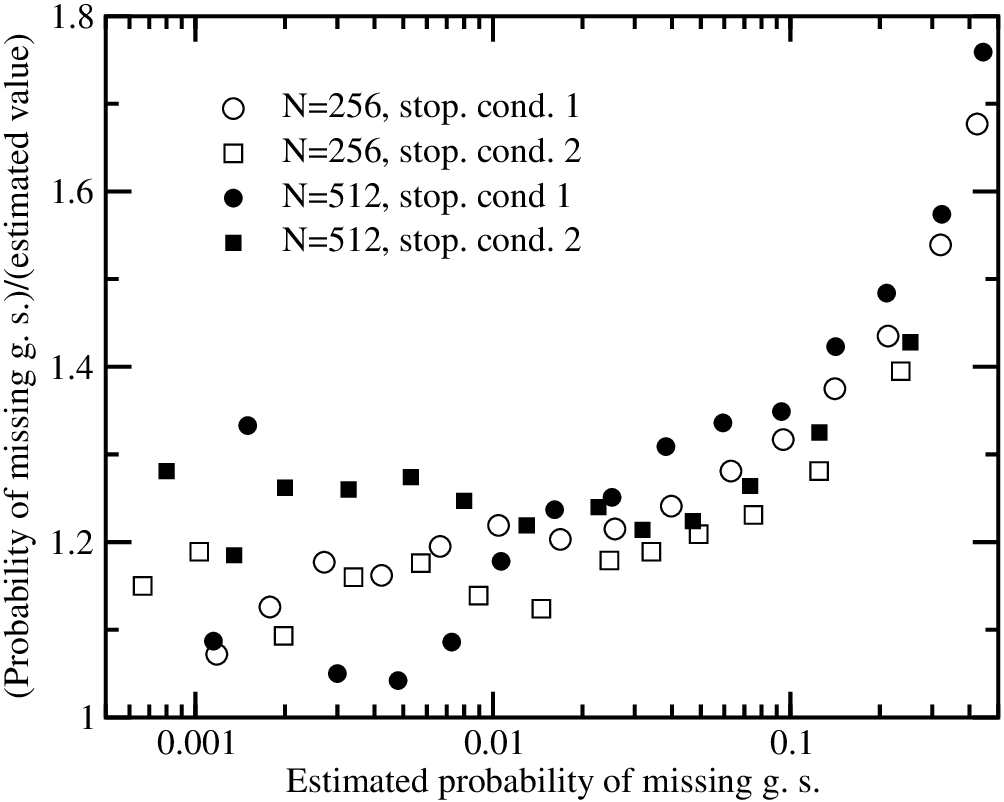}}
\caption{Ratio of the true and the estimated frequency of missing the ground
state with different stopping conditions for samples of $N=256$ and $N=512$
SK systems. See text for details.}
\label{fig:errest}
\end{center} 
\end{figure}

We calculated ground states of large samples between sizes of
$N=64$ and $N=2048$ spins. Parameters of stopping condiditions,
sample sizes, and the estimated probabilities of missing the ground state
for instances with the Gaussian interaction
are shown in Table 1. For $N=128$,  $N=256$ and $N=512$
we made another complete series of calculations, which were longer, and
even more reliable than the first one. We chose $n_{smax}$ very large,
increased $n_{smin}$ to 2000 for $N=128$ and $N=256$, and to 4000 for
$N=512$, and we applied $M_{req}=80$, $M_{req}=50$ and $M_{req}=50$ for
$N=128$, $N=256$ and $512$, respectively.
For N=$128$ all 500000 results agreed, while for $N=256$
and for $N=512$ the longer series improved 6 (0.003\%) and 4 (0.01\%)
cases, respectively, which is a marginal fraction.
For $N=256$ and $N=512$ we derived what energies we would have
ended up if we had applied stopping conditions of type 1 and type 2 with a variety
of $n_s$ (between 1 and 194 for N=$256$ and between 4 and 880 for $N=512$)
and $M_{req}$ (between 2 and 29) values, respectively. As we have every reason
to believe that we know the true ground states of these systems, except for
may be a negligible fraction, we know in each case how often the
states accepted were not the ground states.
Then following
the recipe we proposed above, we estimated the error rates by finding out
how frequently repeated runs would have failed to find the same or a better
energy with the same stopping conditions. Fig.~\ref{fig:errest} shows the factors
we have to multiply the estimated rate of missing the ground state
to get the true error rate in case of the different stopping conditions
considered. The figure shows that we do not underestimate very much the
true error rates. For rates less than about 10\%, this factor is about
$1.2$. It would probably steadily decrease with the error rate,
but our samples are
not large enough to tell how fast. It decreases much slower
than it would if all samples of the same size were equally hard
(in that case the difference between the true and estimated rate would
be about the square of the rate, so for an error rate of $\delta$ the factor shown
on the figure would be about $1+\delta$). Therefore, the error rates shown
in Table \ref{tab:detail} must be very good order of magitude estimates.
Even if the true rates were larger by a factor of 2 or 3 instead of the about 20\% we
believe, it would have a very small effect on the average ground
state energies and on the width of the ground state energy distribution.
Nevertheless, the argument presented here can not be taken as a
strict proof of reliability of the method. To find out how much we
underestimate error rate,
we supposed that the long enough calculations were really
very reliable, i.~e.\  when we see the error rate going to zero, it really does so.
However, if the true ground states of
a finite proportion of the instances were so difficult
for the algorithm to find, that they would not even start appearing
in any calculation of a reasonable length, the assumption would not hold.
We have no reason to believe that the present spin glass
problem were so pathological, therefore we think that the circumstancial
evidence for the reliability of the algorithm is strong enough.

\begin{figure}[ht]
\begin{center}
\scalebox{0.55}{\includegraphics{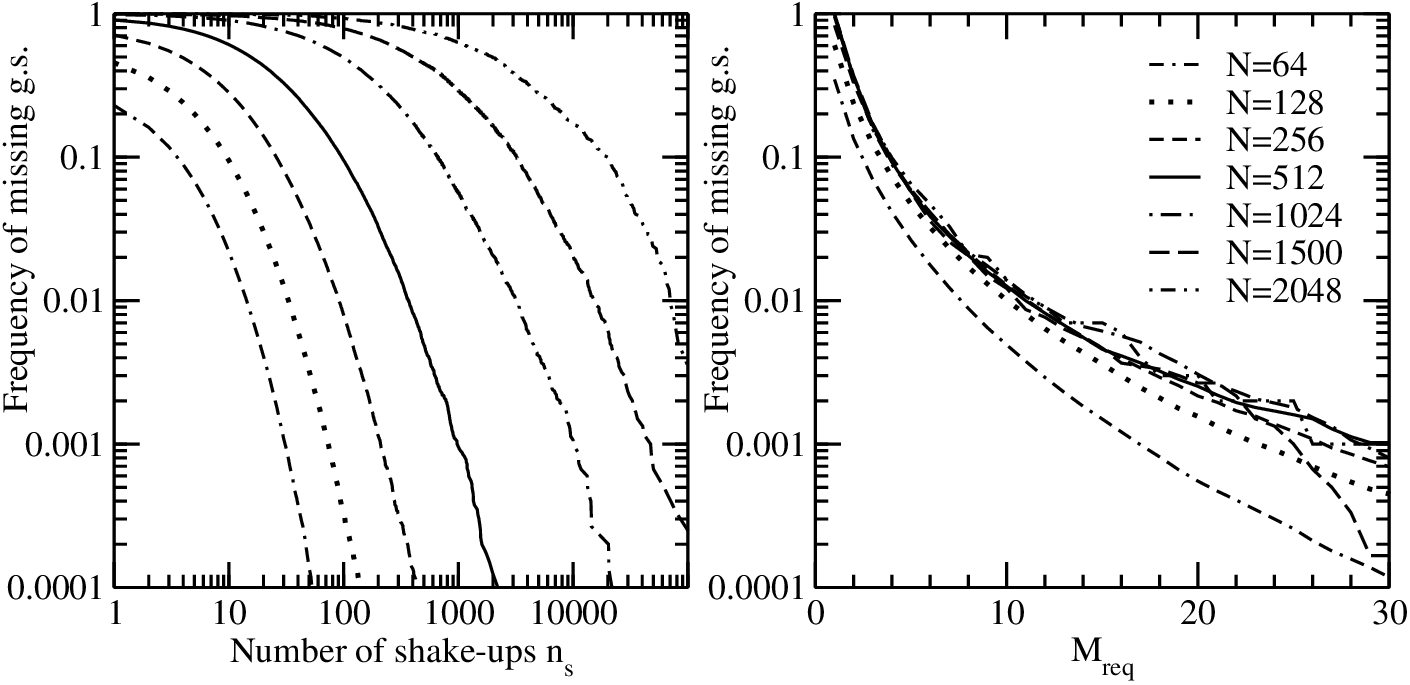}}
\caption{Frequency of not finding the ground state after a fixed
number of shake-ups $n_s$ (stopping condition type 1), and
after the current best state state has been found $M_{req}$ times
(stopping condition type 2).}
\label{fig:errstc}
\end{center} 
\end{figure}

In Fig.~\ref{fig:errstc} we show estimated error rates for different system sizes
with type 1 stopping condition
as the function of the number of shake-ups $n_s$ and with type 2 stopping
condition as the function of the number of times $M_{req}$  we require to
find the lowest state before accepting it as the ground state. We can see that
the number of shake-ups
we need to reach the same reliability grows pretty fast with
$N$. Therefore, making reliable large scale calculations for systems much
larger that $N=2000$ is fairly hopeless with the present algorithm.
It is interesting to note that with the type 2 stopping condition
the reliability of the results with a given  $M_{req}$ does not
seem to depend on the system size $N$ if $N$ is large enough.
If this is not an accident, it may be a further indication
that the results are reliable even for the largest system sizes.

\begin{figure}[ht]
\begin{center}
\scalebox{0.55}{\includegraphics{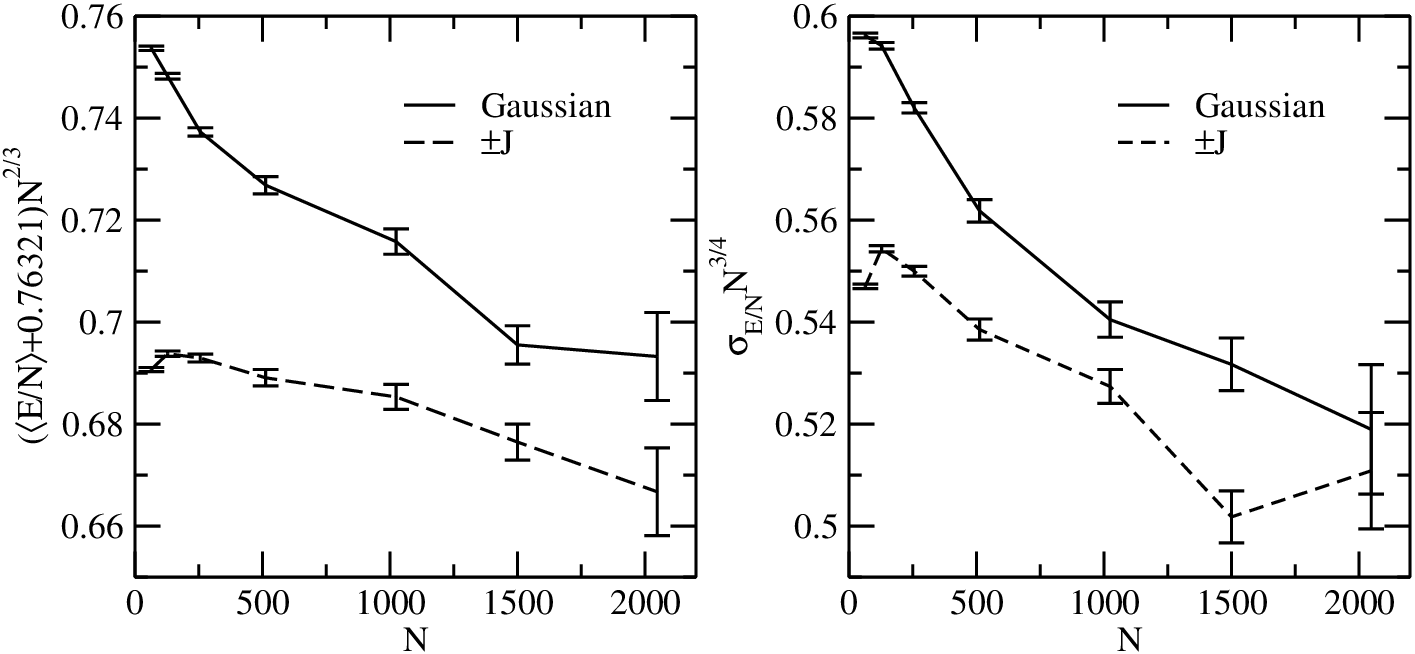}}
\caption{The dependence of the average energy per spin and the width of
its distribution on the size of the system. From the energy the known
asymptotic value is subsctracted, and the functions are multiplied
by factors such that they should behave as constants.}
\label{fig:avsig}
\end{center} 
\end{figure}

Now we discuss shortly some of our results on the behaviour of the
average energy per spin $\langle E(N)/N \rangle$, and the
standard deviation of its distribution $\sigma_{E/N}$. A more detailed
analysis and a discussion of the distribution functions will be
published in a separate paper. The energy per spin
is expected to converge to the asymptotic value with an exponent
$\omega$. This exponent for scaling corrections is not known analytically.
Near $T_c$ a value of $2/3$ has been derived \cite{PaRiSl}, and
numerical studies on smaller systems have been compatible with this value
for the behaviour of the ground state energy as well \cite{BoKrMa, PalasPh,
PalasSK, BoeSK}. For the asymptotic value of $\langle E/N\rangle$ exact result is available
\cite{Par,CrRi}. We substracted this value of -0.76321 from our results for
$\langle E/N\rangle$, and multiplied it with $N^{2/3}$. The result as a function of $N$ is
shown in Fig.~\ref{fig:avsig}, for both the Gaussian and the $\pm J$ case. We would expect to
see horizontal lines if $\omega=2/3$. The energy clearly converges faster
to the asymptotic value than expected, which means that either $\omega$ is
larger than $2/3$, or even with these system sizes we are still not in
the asymptotic region, and subleading corrections are important.
Some deviation has been noted in Ref.~\cite{PalasSK} as well.

The width of the energy per spin distribution converges to zero as
$N^{-\rho}$. The exponent $\rho$ is also unknown analytically.
Ref.~\cite{Kondor} predicted $\rho=5/6$. Since then,
numerical results \cite{CaMaPaPa, BoKrMa, PalasSK, BoeSK} and
qualitative arguments \cite{BoKrMa,AsMoYo} indicated the smaller
value of $\rho=3/4$. We show our results multiplied by $N^{3/4}$ in
Fig.~\ref{fig:avsig}. Convergence is again faster than expected if $\rho=3/4$,
which again means either a larger $\rho$ value, or important subleading
corrections. Boettcher, whose results also started to show deviations
from the expected behaviour \cite{BoeSK}, argued in favour of the
latter possibility. He showed that $\sigma_{E/N}N^{3/4}$ as a function
of $N^{-1/4}$ can well be approximated with a parabola, which
indicates corrections in powers of $N^{-1/4}$. This is actually true for
our results with larger systems and larger samples as well, therefore
$\rho=3/4$ with further corrections is a possibility. However,
$\sigma_{E/N}N^{5/6}$ as a function of $N^{-1/6}$ may also be well
approximated with a parabola, which means that $\rho=5/6$ \cite{Kondor}
is also possible by the same argument. We note that
the parabola indicates not one, but two further non-negligible correction
terms. Actually, if $\rho$ is a simple rational number, for systems larger than
a few hundred spins, our results are most compatible with $\rho=4/5$.
For those larger $N$ values $\sigma_{E/N}N^{4/5}$ is horizontal with
a very good approximation, and no further corrections are needed.

We note that the faster than expected convergence of both the energy and
the width can not be explained by supposing that
our ground states are less reliable than we think. If we do shorter
runs and miss more ground states we get higher values
for both the average and the standard deviation.
For the average this is trivial, and for the standard deviation
it is not surprising either that extra randomness will tend to increase
the spread. As we surely make errors more often for larger systems
than for smaller ones, those errors would make the convergence of
the quantities slower. 

\section{Conclusion}

We have demonstrated in the present paper that hysteretic optimization
is very well suited for finding ground states of Sherrington--Kirkpatrick
spin glasses. The algorithm makes it possible to handle large samples
of systems up to sizes of about 2000 with affordable computational
effort. The method we applied to estimate the proportion of true ground states
missed may be useful also for other algorithms based on
repeating many times the same procedure containing stochastic elements.
Our results show that
the average ground state energy and the width of the ground state energy
distribution converges faster than expected, which may either indicate
larger exponents, or important further correction terms.

\section*{Acknowledgment}

\noindent
The author acknowledges the support of Hungarian grants OTKA T037212 and T037991.

\begin{table}[ht]
\begin{center}
\caption{ \label{tab:detail} Details of hysteretic optimization runs
for the SK model with Gaussian interactions.
System size $N$, sample size, the minimum and the maximum number of
shake-ups $n_{smin}$ and $n_{smax}$, respectively, the required
number of finding the lowest state before accepted as ground state
$M_{req}$, the average CPU time (2.8 GHz P4) per instance $\langle t\rangle$ (the actual time
varies in a wide interval) and the estimated probability of missing the ground
state (see text, the actual value is probably higher by about a factor of 1.2).
}

\begin{tabular}{|c|c|c|c|c|c|c|c|c|}
\hline
\hline
$N$&sample&$n_{smin}$&$n_{smax}$&$M_{req}$&$\langle t\rangle$&
error  \\
\hline
   64&1000000&  200&10000&60&0.06s&0 \\
  128& 500000&  200&10000&60&0.41s&0.001\% \\
  256& 200000&  400&10000&40&2.66s&0.004\% \\
  512&  40000& 1500&15000&30&38.6s&0.013\% \\
 1024&  15000& 1500&15000&20&8.17m&0.23\% \\
 1500&   6000& 3000&30000&15&45.7m&0.58\% \\
 2048&   1000&10000&90000&15&417.m&1.10\% \\
\hline
\end{tabular}
\end{center}
\end{table}

\end{document}